# Theoretical reconstruction of Galileo's two-bucket experiment

Sylvio R. Bistafa[a]

University of São Paulo, São Paulo, Brazil

**Abstract**

In the present work, we address the solution of a problem extracted from a historical context, in which Galileo supposedly conducted an experiment to measure the percussion force of a water jet. To this end, the conservation equations of fluid mechanics in unsteady state are employed in the theoretical reconstruction of the experiment. The experimental apparatus consists of a balance, in which a counterweight hangs on to one of its extremities, and two buckets, in the same vertical, hang on to the other extremity. The water jet issuing from an orifice in the bottom of the upper bucket strikes the lower bucket. The objective is to find the jet percussion force on the lower bucket. The result of the analysis revealed that the method proposed by Galileo for the calculation of the jet percussion force is incorrect. The analysis also revealed that the resultant force during the process is practically null, which would make Galileo's account of the major movements of the balance credible, despite his having not identified all the forces acting on the system.

## 1. Introduction

Galileo's *Discourses* is originally divided into four days — as published in the Leiden edition of 1638 —, to which were posthumously added another two days, all written in dialogic form in *Two New Sciences*. The Sixth Day was translated by Stillman

---

[a] e-mail: sbistafa@usp.br

Drake as the *Added Day: On the Force of Percussion* [1], where the specific goal of the interlocutors Salviati, Sagredo and Aproíno is to understand and find a means of measuring the percussion force.

The first experiment about this force discussed by the trio begins when Aproíno narrates to Sagredo an experiment with two buckets conducted by the Academic (Galileo) to investigate the effect of the percussion force. In this experiment (see Fig. 1), the upper bucket is filled with water and has a hole in the bottom. At the beginning of the experiment the orifice is closed, and the balance is in equilibrium. Once the orifice is opened, the water flows to the lower bucket. Initially the balance tilts to the counterweight side, and after the jet hits the lower bucket the equilibrium is reestablished.

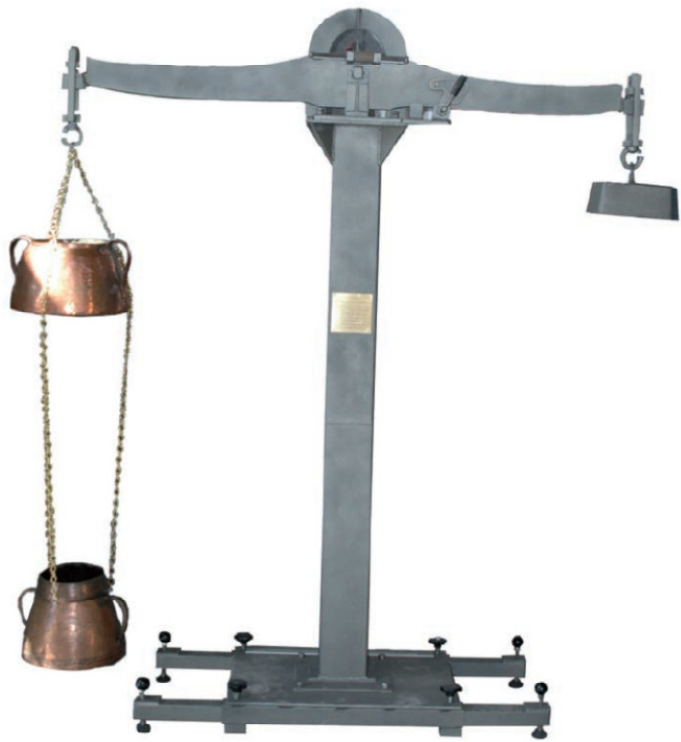

Figure 1. A physical reproduction of Galileo's two-bucket setup. This model is found at the University of Pavia, Italy [From Ref. 2, p. 9].

The present study has the objective of obtaining the forces acting on the balance in unsteady state, since the opening of the orifice in the bottom of the upper bucket until the end of the process, when all the water contained in this bucket has drained to the lower bucket. In the development of the theoretical model, we shall use the conservation equations of fluid mechanics in unsteady state, in the so-called integral form: continuity, in the form of conservation of the volume flux; energy, in the form provided by Torricelli's law; and Newton's 2$^{nd}$ law, best known in fluid mechanics as the linear momentum equation.

## 2. The flow through the orifice and the formation of the water jet

The volume flux $Q_o = Q_o(t)$ through the orifice is given by [3]

$$Q_o = C_d S_o \sqrt{2gh_s}, \tag{1}$$

in which $C_d$ is the discharge coefficient, $S_o$ is the area of the orifice, $h_s = h_s(t)$ is the water height from the orifice up to the free surface of the water in the upper bucket at instant $t$, and $g$ is the gravity. Since Torricelli's law says $V_o = \sqrt{2gh_s}$, then we can write Eq. 1 as $Q_o = C_d S_o V_o$.

The discharge coefficient $C_d$ consists of the product of two other coefficients, namely: the *contraction coefficient* $C_c$, and the *velocity coefficient* $C_v$, such that $C_d = C_c C_v$.

The origin of the contraction coefficient $C_c$ is due to the fact that, as experience shows, the liquid jet cross section at the plane of the orifice $S_o$ continues to contract, until reaching a minimum section, which occurs at a small distance from $S_o$, called *vena contracta*, which is crossed by trajectories that are sensible straight and parallel, in which the velocity is uniform, and the pressure is atmospheric, with the contraction coefficient *theoretically* given by $C_c = \frac{\pi}{\pi+2} \approx 0{,}611$[1]. Torricelli's law refers to the velocity at the vena contracta: in the plane of the orifice, neither the pressure, nor the velocity are uniform, and the velocity is lower than the velocity at the vena contracta.

The velocity given by Torricelli's law $V_o = \sqrt{2gh_s}$ is, however, a theoretical velocity that does not consider the fluid internal viscous forces. Thus, the actual velocity $V_o'$ can be obtained by correcting the theoretical velocity $V_o$ with the velocity coefficient $C_v$, whose value is experimentally obtained. In this way, the actual velocity at the vena contracta is $V_o'$, and given by $V_o' = C_v V_o = C_v\sqrt{2gh_s}$. From this, appears the expression for the volume flux through the orifice as $Q_o = C_c C_v S_o \sqrt{2gh_s} = C_d S_o \sqrt{2gh_s}$, where $C_d = C_c C_v$.

Experience also shows that the cross section of the falling water jet continues to contract, assuming a tapered form as shown in Fig. 2. The shape of the jet during descent may be obtained by applying Bernoulli's equation between point $A$, at elevation $z_A$, and point $B$, at elevation $z_B$, in the form

$$\tfrac{1}{2}\rho V_o'^2 + \rho g z_A + p_A = \tfrac{1}{2}\rho V^2(z) + \rho g z_B + p_B, \qquad (2)$$

where $V_o'$ is the velocity at the *vena contracta*, whose cross section has a radius $a$, $V(z)$ is the velocity at the section whose radius is $r(z)$, $\rho$ is the density, and $p_A$ and $p_B$ are the absolute pressures at $A$ and $B$, respectively.

---

[1] *Vena Contracta*, Kirk T. McDonald
(http://www.physics.princeton.edu/~mcdonand/examples/vena_contracta.pdf)

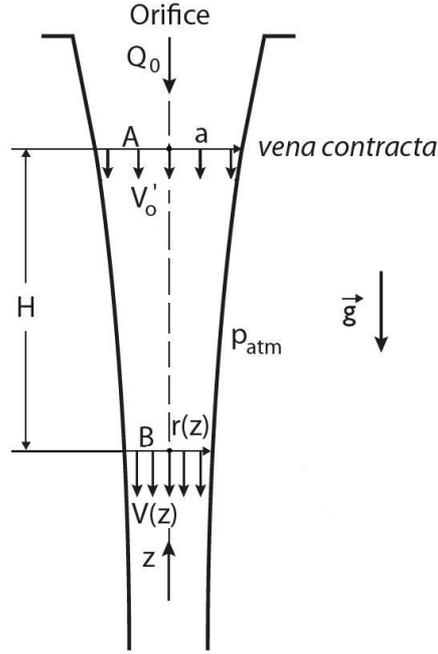

Figure 2. A fluid jet extruded from an orifice accelerates under the influence of gravity. Its shape is influenced both by the gravitational acceleration and surface tension.

The average local curvature $k$ of slender threads may be approximately expressed as $k \approx \frac{1}{r(z)}$. Therefore, the pressures at points $A$ and $B$ may be simply related to the ambient pressure $p_{atm}$ by: $p_A \approx p_{atm} + \frac{\sigma}{a}$, $p_B \approx p_{atm} + \frac{\sigma}{r(z)}$, where $\sigma$ is the surface tension. Substituting these results into Eq. 2, we have

$$\tfrac{1}{2}\rho V_o'^2 + \rho g z_A + p_{atm} + \frac{\sigma}{a} = \tfrac{1}{2}\rho V^2(z) + \rho g z_B + p_{atm} + \frac{\sigma}{r(z)}, \qquad (3)$$

Disregarding the surface tension effects on the shape of the jet, then Eq. 3 may be rewritten as

$$\frac{V(z)}{V_o'} = \left[1 + \frac{2g(z_A - z_B)}{V_o'^2}\right]^{1/2}. \qquad (4)$$

When applying the continuity equation in terms of conservation of the volume flux, for which $\pi a^2 V_o' = \pi r^2 V(z)$, it is possible to rewrite Eq. 4 in the form

$$\frac{r^2}{a^2} = \left[1 + \frac{2g(z_A - z_B)}{V_o'^2}\right]^{-1/2}. \qquad (5)$$

Since $C_c S_o = \pi a^2$, then the area $S$ of any flow cross section is readily obtained from Eq. 5 as

$$S = C_c S_o \left[1 + \frac{2g(z_A - z_B)}{V_o'^2}\right]^{-1/2} = C_c S_o \left(1 + \frac{H}{C_v^2 h_s}\right)^{-1/2}, \qquad (6)$$

where $H = H(t) = z_A - z_B$ is the jet height at instant $t$.

The jet volume $V_j(t)$, at instant $t$, will be given by

$$V_j(t) = \int_0^H S\, dh = C_c S_o \int_0^H \left(1 + \frac{h}{C_v^2 h_s}\right)^{-1/2} dh = 2 S_o C_v^2 C_c h_s(t) \left[\left(1 + \frac{H(t)}{C_v^2 h_s(t)}\right)^{1/2} - 1\right]. \qquad (7)$$

## 3. Theoretical model for the two-bucket experiment in unsteady state

We present next, the linear momentum equation, in the form applicable to the unsteady flow in a control volume $\Omega$, limited by the water contained in the bucket at each instant, as [3]

$$\vec{G} + \vec{R} = \frac{\partial}{\partial t} \int_\Omega \rho \vec{v}\, d\Omega + Q_m V_e \vec{n}_e + Q_m V_s \vec{n}_s, \qquad (8)$$

where $\vec{G}$ is the resultant of the applied body forces (e.g. fluid weight), $\vec{R}$ is the resultant of the forces of contact that act at the walls of control volume; $\frac{\partial}{\partial t}\int_\Omega \rho \vec{v}\, d\Omega$ is the momentum variation inside the control volume $\Omega$; $\vec{v}$ is the fluid velocity, $Q_m V_e \vec{n}_e$ and $Q_m V_s \vec{n}_s$ are the momentum fluxes at the inlet and at the outlet of the control volume, respectively. $Q_m$ is the mass flux, $V_e$ and $V_s$ are the velocities at the inlet and outlet cross sections of the control volume, respectively, and $\vec{n}_e$ and $\vec{n}_s$ are the unit normal vectors at these flow cross sections. The mass flux $Q_m$ can be written as $Q_m = \rho Q$, where $Q$ is the volume flux and $\rho$ is the density of the water (mass per unit of volume).

### 3.1 Resultant force on the upper bucket in unsteady state

The integral $\frac{\partial}{\partial t}\int_{\Omega}\rho\vec{v}\,d\Omega$ in Eq. 8 is the time variation of momentum inside the control volume, which corresponds to a vertical force that decelerates the water mass contained in the upper bucket in its descending motion. During the drainage of the upper bucket, this force acts upward, according to $\vec{e}_z$ (the upward unit vector), decelerating the mass of water in its descending motion. Since, by hypothesis, all the water particles inside $\Omega$ move downward with the same velocity of the free surface $\vec{V}_e$, then, the integral $\frac{\partial}{\partial t}\int_{\Omega}\rho\vec{v}\,d\Omega = \rho\int_0^{h_s}\frac{\partial \vec{V}_e}{\partial t}S_e dh$, in which $S_e$ is the area of the flow cross sections in the upper bucket.

The integral $\rho\int_0^{h_s}\frac{\partial \vec{V}_e}{\partial t}S_e dh = \rho S_e h_s \frac{\partial \vec{V}_e}{\partial t}$, in which $h_s = h_s(t)$ is the height of the free surface of the water in upper bucket at instant $t$. In this expression, $\frac{\partial \vec{V}_e}{\partial t}$ is the acceleration to which the water in the upper bucket is subjected to during its descending motion. Then, $\frac{\partial \vec{V}_e}{\partial t} = \frac{\partial V_e}{\partial t}(-\vec{e}_z)$, and, since $V_e = C_d \frac{S_o}{S_e}\sqrt{2gh_s}$, results in $\frac{\partial \vec{V}_e}{\partial t} = C_d \frac{S_o}{S_e}\frac{g}{\sqrt{2gh_s}}\frac{\partial h_s}{\partial t}(-\vec{e}_z)$.

But, from continuity $\frac{\partial h_s}{\partial t} = -C_d \frac{S_o}{S_e}\sqrt{2gh_s}$, and then, $\frac{\partial \vec{V}_e}{\partial t} = gC_d^2 S_e \left(\frac{S_o}{S_e}\right)^2 h_s(\vec{e}_z)$.

Then, finally, $\frac{\partial}{\partial t}\int_{\Omega}\rho\vec{v}\,d\Omega$ can be written as

$$\frac{\partial}{\partial t}\int_{\Omega}\rho\vec{v}\,d\Omega = \rho g C_d^2 S_e \left(\frac{S_o}{S_e}\right)^2 h_s(\vec{e}_z), \qquad (9)$$

for the upper bucket.

Substituting Eq. 9 into Eq. 8, and considering that in Eq. 8 the momentum flux $Q_m V_e \vec{n}_e = 2\rho g C_d^2 \frac{S_0^2}{S_e} h_s(\vec{e}_z)$ and the momentum flux $Q_m V_s \vec{n}_s = 2\rho g C_d C_v S_0 h_s(-\vec{e}_z)$, results in

$$\vec{R}_{bs} = G_{lbs}(\vec{e}_z) + \rho g C_d^2 S_e \left(\frac{S_0}{S_e}\right)^2 h_s(\vec{e}_z) + 2\rho g C_d^2 \frac{S_0^2}{S_e} h_s(\vec{e}_z) + 2\rho g C_d C_v S_0 h_s(-\vec{e}_z),$$

(10)

in which $G_{lbs}$ is the weight of the water contained in the upper bucket. $\vec{R}_{bs}$, as given by Eq. 10, is the resultant force acting on the water body contained in the upper bucket in unsteady state.

### 3.2 Resultant force on the lower bucket in unsteady state

As far as the lower bucket is concerned, the velocity of the jet as it strikes the bottom of this bucket changes its direction from axial to the radial direction; then, $\vec{V}_e = V_e(-\vec{e}_z) = 0$. Therefore, for the lower bucket $\frac{\partial}{\partial t}\int_\Omega \rho \vec{v}\, d\Omega = \rho \int_\Omega \frac{\partial \vec{V}_e}{\partial t} d\Omega = 0$.

The momentum flux of the jet falling on the free surface of the lower bucket is given by $Q_m V_e \vec{n}_e = \rho Q V_e(\vec{e}_z)$, where $Q = Q(t)$ is the volume flux that enters the lower bucket, and that varies with the time in the unsteady state.

For a fixed control volume that incorporates the volume of the water jet, the continuity equation, in unsteady state, for an incompressible fluid may be written as

$$Q = Q_o - \frac{\partial}{\partial t}\int_{V_j} dV_j = Q_o - \frac{\partial V_j}{\partial t},$$

(11)

in which $Q_o = C_d S_0 \sqrt{2gh_s}$ is the volume flux through the orifice and $V_j$ is the volume of the jet at each instant, as given by Eq. 7.

Applying Eq. 7 in the assessment of $\frac{\partial V_j}{\partial t}$ we have that

$$\frac{\partial V_j}{\partial t} = \frac{\partial}{\partial t}\left\{2S_o C_v{}^2 C_c h_s\left[\left(1+\frac{H}{C_v{}^2 h_s}\right)^{1/2}-1\right]\right\}. \tag{12}$$

Once the derivative indicated in Eq. 12 has been evaluated, and substituting the resulting expression into Eq. 11, gives

$$Q = Q_o\left\{1 - 2\frac{S_o}{S_e}C_v{}^2 C_c\left[\frac{1}{2C_v{}^2}\left(\frac{H}{h_s}-1\right)\left(1+\frac{H}{C_v{}^2 h_s}\right)^{-1/2} - \left(1+\frac{H}{C_v{}^2 h_s}\right)^{1/2} + 1\right]\right\}. \tag{13}$$

Then, the momentum flux of the jet that falls on the free surface of the lower bucket $Q_m V_e \vec{n}_e = \rho Q V_e(\vec{e}_z)$, with $V_e = \frac{Q}{S}$, in which $S$ is the area of the jet cross section when falling on the free surface. This area is given by Eq. 6, repeated here as $S = C_c S_o\left(1+\frac{H}{C_v{}^2 h_s}\right)^{-1/2}$; and, therefore $Q_m V_e \vec{n}_e = \rho\frac{Q^2}{S}(\vec{e}_z) = \rho Q^2 (C_c S_o)^{-1}\left(1+\frac{H}{C_v{}^2 h_s}\right)^{1/2}(\vec{e}_z)$.

On the other hand, the momentum flux of the lower bucket free surface will be given by $Q_m V_s \vec{n}_s = \rho\frac{1}{S_e}Q^2(\vec{e}_z)$, where $S_e$ is the free surface area of the lower bucket.

Then, from these results, we may write Eq. 8 for the lower bucket as

$$\vec{R}_{bi} = G_{lbi}(\vec{e}_z) + \rho Q^2 (C_c S_o)^{-1}\left(1+\frac{H}{C_v{}^2 h_s}\right)^{1/2}(\vec{e}_z) + \rho\frac{1}{S_e}Q^2(\vec{e}_z), \tag{14}$$

in which $G_{lbi}$ is the weight of the water contained in the lower bucket, with $Q$ given by Eq. (13). $\vec{R}_{bi}$, as given by Eq. 14, is the resultant force acting on the water body contained in the lower bucket in unsteady state.

### 3.3 Resultant force on the balance in unsteady state

The resultant force on the balance $\vec{R}_b$ will be given by the sum of $-\vec{R}_{bs}$ (Eq. 10) and $-\vec{R}_{bi}$ (Eq. 14). By considering that the weight of the water in the system $G_l$ may be written as $G_l = G_j(t) + G_{lbs}(t) + G_{lbi}(t)$, or $G_l - G_j(t) = G_{lbs}(t) + G_{lbi}(t)$, where $G_j(t) = \rho g V_j(t)$ is the weight of the water jet suspended in the air between the two buckets at instant $t$, the resultant force on the balance will be given by

$$\vec{R}_b = \left\{ G_l - G_j(t) + \rho g C_d^2 S_e \left(\frac{S_o}{S_e}\right)^2 h_s + 2\rho g C_d^2 \frac{S_o^2}{S_e} h_s - 2\rho g C_d C_v S_o h_s + \right.$$

$$\left. \rho Q^2 (C_c S_o)^{-1} \left(1 + \frac{H}{C_v^2 h_s}\right)^{1/2} + \rho \frac{1}{S_e} Q^2 \right\} (-\vec{e}_z). \tag{15}$$

The division of Eq. 15 by the weight of the water contained in the system $G_l$, and given by $G_l = \rho g L S_e$, where $L$ is the height of the water in the upper bucket in the beginning of its drainage, gives the dimensionless form of this equation as

$$\frac{\vec{R}_b}{G_l} = \left\{ \frac{G_l}{\rho g L S_e} - 2 C_d C_v R C^{-1} H_s \left[\left(1 + \frac{H}{C_v^2 H_s}\right)^{1/2} - 1\right] + C_d^2 R C^{-2} H_s + 2 C_d^2 R C^{-2} H_s - \right.$$

$$\left. 2 C_d C_v R C^{-1} H_s + \frac{Q^2}{g L S_e} (C_c S_o)^{-1} \left(1 + \frac{H}{C_v^2 H_s}\right)^{1/2} + \frac{Q^2}{g L S_e^2} \right\} (-\vec{e}_z). \tag{16}$$

Calling $\frac{h_s}{L}$ by $H_s$, $\frac{H}{L}$ by $H$, and $\frac{S_e}{S_o}$ by $RC$, where $RC$ is the *contraction ratio*, then Eq. 16 may be rewritten in the most general form as

$$\frac{\vec{R}_b}{G_l} = \left\{ \frac{G_l}{\rho g L S_e} - 2 C_d C_v R C^{-1} H_s \left[\left(1 + \frac{H}{C_v^2 H_s}\right)^{1/2} - 1\right] + C_d^2 R C^{-2} H_s + 2 C_d^2 R C^{-2} H_s - \right.$$

$$\left. 2 C_d C_v R C^{-1} H_s + \frac{Q^2}{g L S_e} (C_c S_o)^{-1} \left(1 + \frac{H}{C_v^2 H_s}\right)^{1/2} + \frac{Q^2}{g L S_e^2} \right\} (-\vec{e}_z). \tag{17}$$

Let us rename the terms that appear in Eq. 17, by calling: $\boldsymbol{A} = \frac{G_l}{\rho g L S_e} = 1$, the relative weight of the water contained in the system; $\boldsymbol{B} = 2 C_d C_v R C^{-1} H_s \left[\left(1 + \frac{H}{C_v^2 H_s}\right)^{1/2} - 1\right]$, the relative weight of the water suspended in the air between the orifice and the free surface of the lower bucket; $\boldsymbol{C} = C_d^2 R C^{-2} H_s$, the relative variation of momentum in the upper bucket; $\boldsymbol{D} = 2 C_d^2 R C^{-2} H_s$, the relative momentum flux at the upper bucket free surface; $\boldsymbol{E} = 2 C_d^2 C_c^{-1} R C^{-1} H_s$, the relative momentum flux at the orifice; $\boldsymbol{F} = \frac{Q^2}{g L S_e} (C_c S_o)^{-1} \left(1 + \frac{H}{C_v^2 H_s}\right)^{1/2}$, the relative momentum flux of the jet that falls into the free surface of the lower bucket; and $G = \frac{Q^2}{gLS_e^2}$, the relative momentum flux at the free surface of the lower bucket.

By substituting $Q$, given by Eq. 13, into the expressions for $F$ and $G$, the result is

$$F = 2C_d^2 C_c^{-1} RC^{-1} H_s \left\{ 1 - 2C_v^2 C_c RC^{-1} \left[ \frac{1}{2C_v^2} \left( \frac{H}{H_s} - 1 \right) \left( 1 + \frac{H}{C_v^2 H_s} \right)^{-1/2} - \left( 1 + \frac{H}{C_v^2 H_s} \right)^{1/2} + 1 \right] \right\}^2 \left( 1 + \frac{H}{C_v^2 H_s} \right)^{1/2}. \tag{18}$$

$$G = 2C_d^2 RC^{-2} H_s \left\{ 1 - 2C_v^2 C_c RC^{-1} \left[ \frac{1}{2C_v^2} \left( \frac{H}{H_s} - 1 \right) \left( 1 + \frac{H}{C_v^2 H_s} \right)^{-1/2} - \left( 1 + \frac{H}{C_v^2 H_s} \right)^{1/2} + 1 \right] \right\}^2. \tag{19}$$

Finally, we may write Eq. 17 in a more compact form, as

$$\frac{\vec{R}_b}{\rho g L S_e} = (A - B + C + D - E + F + G)(-\vec{e}_z). \tag{20}$$

Figure 3 highlights the forces that appear in Eq. 20.

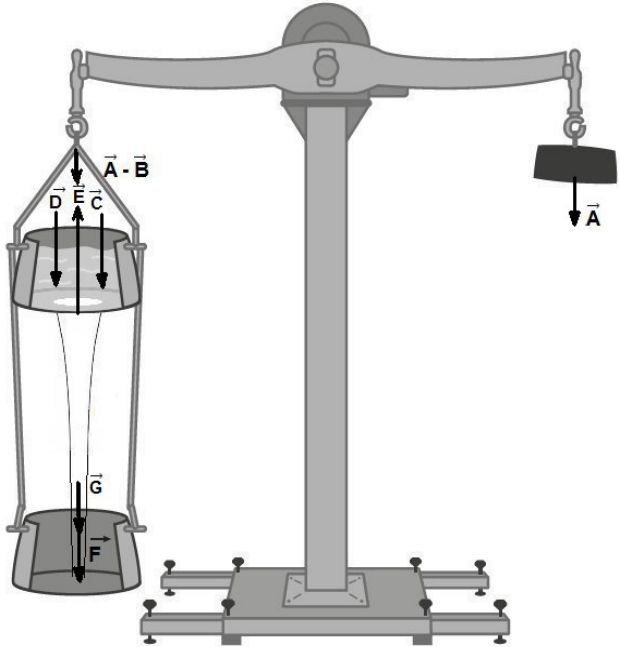

Figure 3. Relative forces acting on the balance: weight of the water in the system – **A**, weight of the jet – **B**, variation of momentum in the upper bucket – **C**, momentum flux at free surface of the upper bucket – **D**, reaction force in the upper bucket – **E**, percussion force in the lower bucket – **F**, momentum flux at the free surface of the lower bucket – **G**.

To numerically evaluate Eq. 20, we need, now, an expression that relates the time elapsed to the height of the water free surface in the upper bucket. For the determination of this elapsed time, we shall write the continuity equation for the upper bucket in the form $Q(t) = -S_e \frac{dh_s}{dt}$, where $Q(t) = C_d S_o \sqrt{2gh_s}$; and hence

$$\frac{dh_s}{dt} = -C_d \frac{S_o}{S_e}\sqrt{2gh_s}. \tag{21}$$

Upon integration, Eq. 21 yields

$$t = \frac{S_e}{C_d S_o}\sqrt{\frac{2}{g}}(\sqrt{h_o} - \sqrt{h_s}), \tag{22}$$

where $h_o = h(t=0) = L$.

By writing $\frac{S_e}{S_o} = RC$, $\frac{h_o}{L} = 1$, and $\frac{h_s}{L} = H_s$, then Eq. 22 is transformed into

$$t = \frac{RC}{C_d}\sqrt{2\frac{L}{g}}(1 - \sqrt{H_s}), \tag{23}$$

valid for $0 \leq H_s \leq 1$. This expression will give the time elapsed since the opening of the orifice, until the instant when the water height in the upper bucket reaches the value $H_s$.

Galileo, supposedly used in his experiment elements with the following dimensions [1]: distance between the bottoms of the buckets $H_i$ equal to 1.35 m (two *braccia*; one *braccio* ~ 67 cm), and orifice with diameter[2] equal to 0.03 m. The diameter of the buckets and the water height in the upper bucket $L$ are not narrated, and, therefore, it was assumed that both are equal to 0.3 m. For these dimensions, the volume of water contained in the system is 21.2 liters, with a mass of 21.2 kg and weight of 208 N, approximately. This is the weight of the water contained in the system, called here as $G_l$. With these numerical values, we have $RC = \frac{S_e}{S_o} = \left(\frac{0.3\ m}{0.03\ m}\right)^2 = 100$.

---

[2] Galileo indicates that "[…] The bottom of the upper bucket had been pierced by a hole the size of an egg or a little smaller" [1]. There is no doubt that nowadays, the eggs are greater than their homologues in Galileo's time. A search over the Internet revealed that the average diameter of a chicken egg in its larger cross section is around 4.25 cm. A diameter of this magnitude would drain the bucket very quickly, not allowing an adequate observation of the movement of the balance. For these reasons, it was decided to adopt an orifice diameter of 3 cm.

The following values for the flow coefficients were adopted in the calculations as representative of the process: $C_c = 0.63$, $C_v = 0.97$, $C_d = 0.61$ [3].

Figure 4 presents the forces acting on the balance, obtained using Eq. 20 and 23, for $RC = 100$, since the opening of the orifice in the bottom of the upper bucket, until its complete drainage. In this figure, forces with positive values tend to move the balance toward the buckets' side, and forces with negative values tend to move the balance toward the counterweight side.

It is clearly seen in Fig. 4 that the forces that dominate the process are: the weight of the jet – **B**, the reaction force on the upper bucket – **E**, and the percussion force on the lower bucket – **F**. The force generated by the momentum variation in the upper bucket – **C**, the force generated by the momentum flux at the free surface of the upper bucket – **D**, and the force generated by the momentum flux at the free surface of the lower bucket – **G** are practically null during the entire drainage process. Fig. 4 also indicates that, after the first percussion of the jet in the lower bucket, the resultant force is practically null during the upper bucket drainage. Thus, accordingly, the balance, that was inclined toward the counterweight side after the opening of the orifice, will tend to return to the equilibrium position, remaining, however, still a little inclined toward the buckets side during the process.

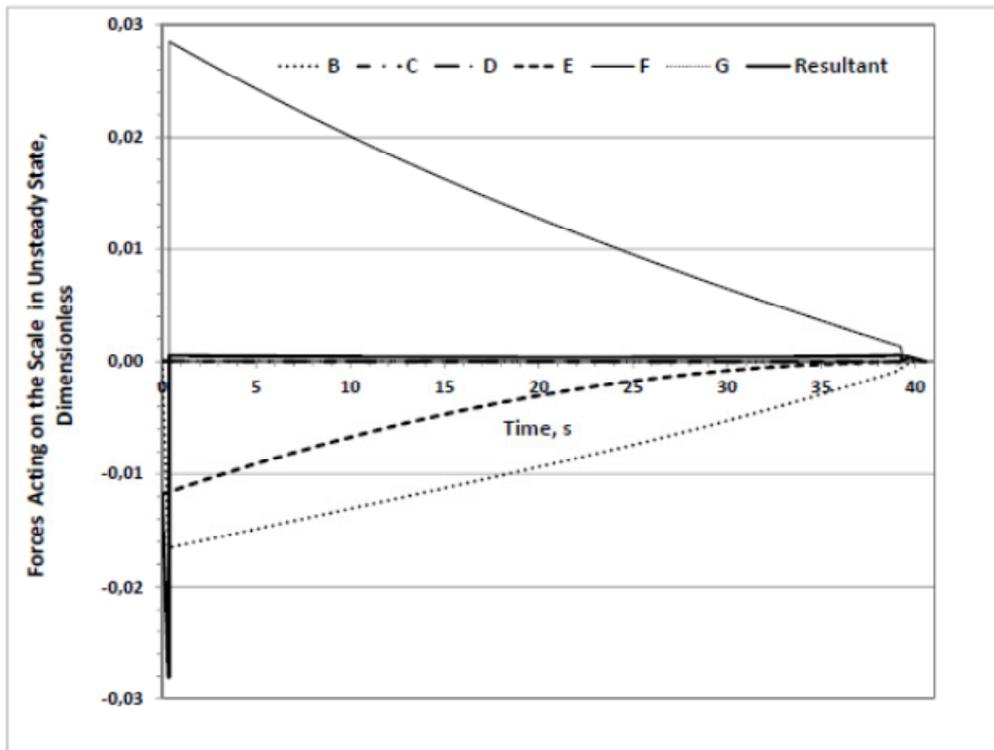

Figure 4. Forces acting on the balance, since the opening of the orifice in the upper bucket, until the lower bucket is completely filled, for $RC = 100$. Weight of the jet – **B**, variation of momentum in the upper bucket – **C**, momentum flux at free surface of the upper bucket – **D**, reaction force on the upper bucket – **E**, percussion force on the lower bucket – **F**, momentum flux at the free surface of the lower bucket – **G**, and the **Resultant Force**.

## 4. Discussion

For Galileo, the percussion force would be equal to the weight of the jet that is suspended in the air between the waters in the two buckets [1]. However, Fig. 5 shows that the percussion force has a different behavior from the weight of the jet, with a value always greater during the upper bucket drainage.

Aproíno, in his talk [1], states that the weight of the jet would be 'certainly' 10 to 12 pounds, although Salviati indicates in his replica that there would be some uncertainty due to the 'difficulty in measuring the amount of the falling water'. Although Aproíno does not mention at which instant of time this value would have been obtained, it may be admitted that it could be at the instant when the jet first strikes the lower bucket. At this instant, the percussion force corresponds to, approximately, 2.8% of the weight of water contained in the system, which gives 1.75 pounds.[3] At this same instant of time, the weight of the jet corresponds, approximately, to 1.7% of the weight of water contained in the system, that is, 1.06 pounds, which is a much lower value than that estimated by Galileo.

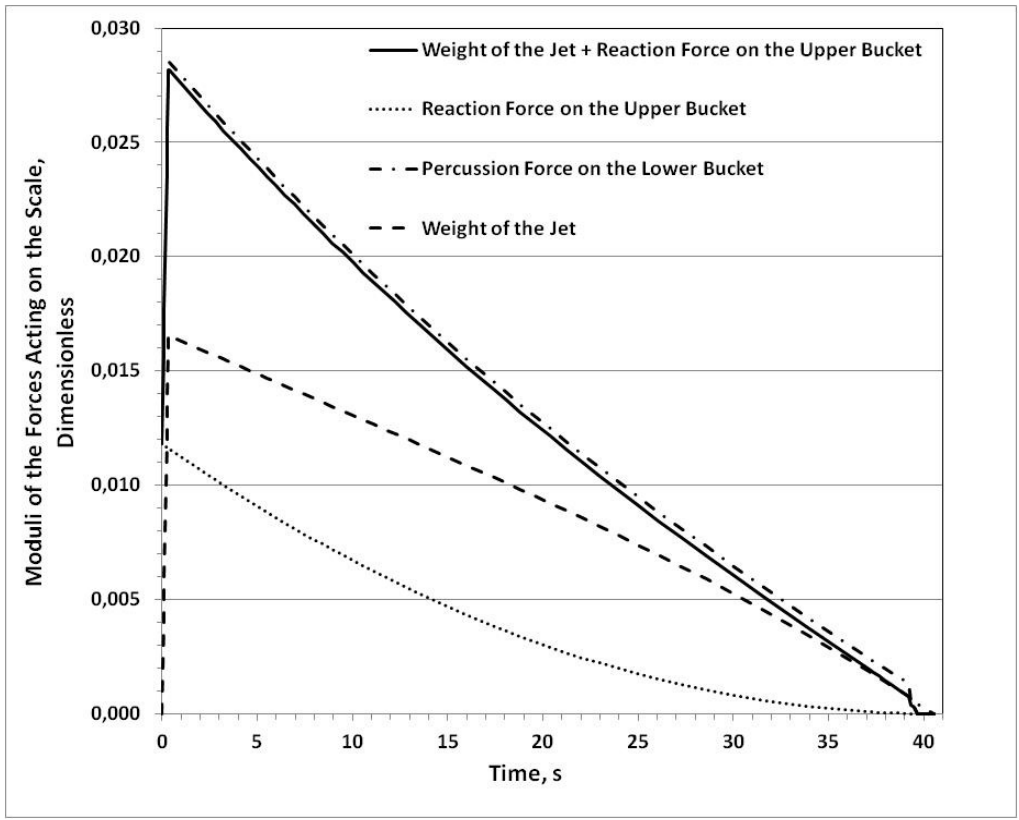

Figure 5. Moduli of the forces acting on the balance, for $RC = 100$.

Figure 5 also indicates that during the drainage of the upper bucket, the percussion force is practically equal to the sum of the weight of the jet, plus the reaction force in the upper bucket. The small difference between the percussion force and this sum is practically constant during the entire process, around 0.03% of the weight of the water contained in the system, corresponding to a resultant of only 0.062 N (6.4 grams), approximately, in favor of the percussion force. This resultant will cause the balance to remain a little unbalanced toward the buckets side during the upper bucket drainage.

---

[3] 1 *Tuscan pound* = 0.3395 kg, Ref. [2].

## 5. Conclusions

The analysis made demonstrates that the percussion force in the lower bucket does not correspond to the weight of the jet that is suspended in the air between the waters in the two buckets, upper and lower, as suggested by Galileo. In fact, the percussion force is proportional to the square of the jet velocity, assuming a value always greater than the weight of the jet during the upper bucket drainage.

During the upper bucket drainage, the balance will remain a little unbalanced toward the side of the buckets, but due to the small magnitude of resultant force, with a value practically constant, and around 6.4 grams only during the entire process – which would make the unbalance of the balance described by Galileo small enough to pass unnoticed –, indicates that the report of Galileo could be considered as being credible "[…] but the water had hardly begun to strike against the bottom of the lower bucket when the counterweight ceased to descend, and commenced to rise with very tranquil motion, restoring itself to equilibrium while water was still flowing, and upon reaching equilibrium it balanced and came to rest without passing a hairbreadth beyond." [1].

**References**

[1] Stillman Drake (1989) *Galileo Galilei: Two New Sciences* 2$^{nd}$ edn. (Ontario: Wall & Emerson, Inc.) 281−342 (http://www.spirasolaris.ca/sbb6Added_Day.pdf).

[2] Roberto Vergara Caffarelli (2009) *Galileo Galilei and Motion - A Reconstruction of 50 Years of Experiments and Discoveries* (Jointly published by Springer & Società Italiana di Fisica).

[3] Tufí Mamede Ássy (2004) *Mecânica dos Fluidos: Fundamentos e Aplicações* 2$^{nd}$ edn. (Rio de Janeiro: LTC).